\documentstyle[aps,epsf,prl,twocolumn]{revtex}
\begin{document}
\vspace{0.0cm}
\draft
\title{
Classical statistical mechanics of a few--body interacting spin
model }
\author{F.Borgonovi$^{[a,b,c,\dagger]}$and
F.M.Izrailev$^{[d]}$}
\address{
$^{[a]}$Department of Physics, University of Maryland, College
Park, MD 20742\\
$^{[b]}$Istituto Nazionale di Fisica della
Materia, Unit\`a di Milano, via Celoria 16, 22100 Milano, Italy\\
$^{[c]}$Istituto Nazionale di Fisica  Nucleare,
Unit\`a di Pavia, via Bassi 6, 27100 Pavia, Italy\\
$^{[d]}$Instituto de F\'{\i}sica, Universidad Aut\'onoma de
Puebla, Apartado Postal J-48, Puebla, 72570 M\'exico
\\
}
\date{\today}
\maketitle

\begin{abstract}
We study the emergence of  Boltzmann's  law for the ``single
particle energy distribution" in a closed system of interacting
classical spins. It is shown that for a large number of particles
 Boltzmann's law may occur, even if the interaction is very
strong.  Specific attention is paid to classical analogs of the
average shape of quantum eigenstates and "local density of states",
which are very important in quantum chaology. Analytical
predictions are then compared with numerical data.
\end{abstract}
\pacs{PACS numbers: 05.45.-a}


\section{Introduction}
Chaotic properties of few degrees of freedom systems have
attracted much attention during the last years. The knowledge and
classification of chaotic dynamical systems is now extremely
accurate, if compared with the status of art at the beginning of
the century. On the other hand, conventional statistical mechanics
was born long before the chaology of dynamical systems. It is well
known that, in contrast to chaotic dynamical systems with few
degrees of freedom, for the onset of statistical equilibrium in
systems with an infinitely large number of particles
neither the non-linearity nor the interaction between
particles is needed. Indeed, in such systems the thermodynamic limit
(infinite number of particles, $N\rightarrow
\infty $ ) leads to the statistical behavior of a system since any weak
interaction (with an environment or between particles) gives rise
to mixing properties and onset of chaos.

Although statistical mechanics has obtained undoubtedly results in
many different fields, fundamental questions about statistical
description of systems with finite number of particles are still
open\cite{schuster}. It is clear that in such systems the interparticle
interaction is crucial, however, its role in producing chaos or
ergodicity is still not understood. Due to computing difficulties
and  lack of theoretical studies, many--body chaotic systems
have been scarcely investigated and little is known about them.

Recently, a novel approach to quantum isolated systems of  finite
number of interacting particles has been developed
\cite{FIC96,FI97,FI97a,FI99,I99}. The goal of this approach
is a direct relation between the average shape of exact
eigenstates ($F-$function), and the distribution of occupation
numbers $n_s$ of single-particle levels. This relation shows that
there is no need to know exactly the eigenstates, instead, if
these eigenstates are chaotic (random superposition of a very
large number of components of basis states), the $F-$ function
absorbs the statistical effects of interaction between particles and
determines the form of the $n_s-$distribution. The approach has
been mainly developed for a model with completely random two-body
interaction between a finite number of Fermi-particles, however, it
can be also applied to dynamical models with chaotic behavior.

One of few dynamical models studied with the use of this approach
is a system of two interacting spins \cite{BGIC98,BGI98}. The most
interesting result obtained numerically in Ref.\cite{BGIC98}, is
that the distribution of occupation numbers (even for two
interacting particles) in the chaotic region can be described in
the form of the standard Bose-Einstein (BE) distribution (only
symmetric states have been considered which results in the
BE statistics). It was shown that the statistical effects of the
interaction lead to an increase of temperature which is related to
the BE distribution. Also, it was discovered that the canonical
distribution is recovered if one randomize the {\it non-zero
elements} of the interaction $V$ keeping the dynamical constrains of
the model. This means that random interaction plays the  role of a
heat bath and allows to use statistical and thermodynamical
description for isolated systems.

In order to extend the approach of
Refs.\cite{FIC96,FI97,FI97a,FI99,I99} to classical dynamical
systems with a large number of particles, we introduce below the
model of $N$ interacting spins in one-dimension. Due to the
well-defined classical limit, it is of the first interest to
explore similar problems in the classical counterpart. The problem
of a quantum-classical correspondence for chaotic systems with a large
number of interacting particles seems to us extremely important in
view of many physical applications. In this paper we mainly
concentrate on the occurrence of  Boltzmann's law and on the
classical counterparts of quantum local density of states and
shape of eigenfunctions. We deserve for future work
\cite{BI99} the investigation of the analogous Fermi-Dirac or
Bose--Einstein statistics in a quantized version of such a model.

Our investigation is complementary to the  approach based on the
so-called {\it dynamical} temperature \cite{R97,BKP97,B98},
introduced for the study of the statistical
properties of few degrees of freedom
classical  models with chaotic behavior.
Specifically, our interest is in the
notion of {\it statistical} temperature associated with
Boltzmann's   distribution (if any) for  single
particle energy. The
relation between statistical and dynamical temperatures
represents an open and interesting question.

\section{The model of interacting spins}

Our purpose is to investigate a few--body dynamical system from
the statistical mechanics point of view. Following previous works
on two--particle spin problem\cite{BGI98,BGIC98}, we consider the
Hamiltonian:
\begin{equation}
H = B\sum_{i=1}^{N}S_i^z + J\sum_{i=1}^{N} \vec{S}_i \cdot
\vec{S}_{i+1}
\label{hb}
\end{equation}
which is known as the 1-d Heisenberg model in a magnetic field,
see for instance \cite{M93}. Here $\vec S_i$ are spin vectors,
and, for sake of definiteness we take periodic boundary
conditions, $\vec S_1=\vec S_{N+1}$. The classical version is a
solvable model, from the statistical point of view (namely when
the number of particles $N\to \infty $, see \cite{F64}).

Instead, we are here interested in a dynamical approach, in
particular, for a small number of particles,
when the usual statistical approach is at least questionable. In
what follows, we shall consider a more simple version of the
classical Heisenberg model, which is described by the following
Hamiltonian~:
\begin{equation}
\begin{array}{ll}
H = H_0 + V = B \sum_{i=1}^N S_i^z + J \sum_{i=1}^N S_i^y
S_{i+1}^y
\end{array}
\label{hint}
\end{equation}
where $N>2$ is the number of spins in the chain. This we do, in
order to simplify analytical calculations. The equations of motion
can be written in the usual way:~

\begin{equation}
\label{eom}{\frac{{d{\vec S}}}{{dt}}}=\{H,\vec S\}
\end{equation}
where $\{,\}$ are the Poisson brackets (see
\cite{Feingold,glialtri} for similar dynamical models). Constants
of motion are the energy $E$ and the magnitude of the angular
momenta $|\vec S_i|=s$ (the latter  assumed to be the same for
each spin). Without loss of generality, we can put $s=1$.

Contrary to the common view point (spin Hamiltonian plus a
magnetic field as a perturbation) we consider the ``magnetic''
part as an unperturbed Hamiltonian. Indeed, we are interested in
models which can be generally expressed as a sum of single
particle Hamiltonians, this feature not being shared by the
perturbation $V$. Therefore, the Heisenberg kernel ($J\sum
\vec S_i\vec S_{i+1}$) will be considered as a perturbation
between nearest-neighbor spins.

\begin{figure}
\epsfxsize 7cm
\epsfbox{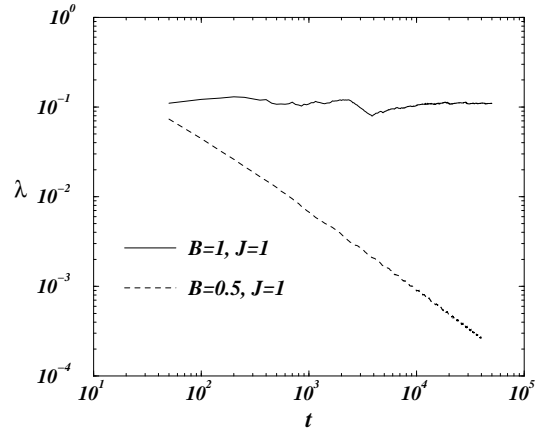}
\narrowtext
\caption{ Maximal Lyapunov exponents as a function of time
for $E=-2$, $J=1$ and different $B=0.5,1$ as indicated in the
picture.  The  system has $N=4$ spins.
}
\label{lyap1}
\end{figure}

The unperturbed Hamiltonian is integrable and the solution of the
unperturbed equations of motion can be written down at glance. It should be
pointed out that the perturbation itself is not chaotic.
Indeed, it can be verified numerically that two close trajectories
diverge only linearly in time when  $B=0$. This means that the
maximal Lyapunov exponent is zero for any choice of the initial
conditions. On the other side the same Lyapunov analysis shows
that the total Hamiltonian, for generic values of the coupling $J$
and for energy values in the middle of the band $(E\sim 0)$ is
chaotic. Maximal Lyapunov exponents, as a function of time for
different $J$ and $B$ values and fixed energy $E$ , are shown in
Fig.\ref{lyap1}.

\begin{figure}
\epsfxsize 7cm
\epsfbox{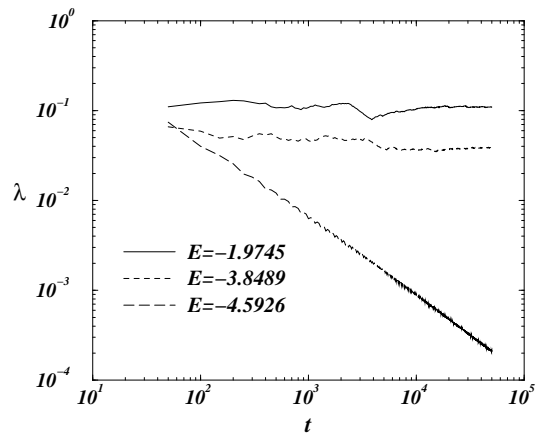}
\narrowtext
\caption{ a) Maximal Lyapunov exponents as a function of time
for $J=1$, $B=1$ and different  energy $E$ as indicated in the
text. Simulations have been done with  $4$ spins.
}
\label{lyap2}
\end{figure}

As one can see, the Lyapunov exponents approach zero when $B$
decreases, or $J$ increases. On the other side, for fixed $J$ and
$B$ different from zero, the region where the maximal Lyapunov
exponent approaches zero is close to the edge of the energy
spectrum (see Fig.\ref{lyap2}). Maximal Lyapunov exponents have
been calculated using the standard recipe \cite{LL92}.

Besides integrability, the unperturbed Hamiltonian $H_0$,
in the limit $N\rightarrow \infty \,$, has good statistical
properties which we are going to exploit in the next Section. It
is, indeed, interesting to compare our results with those obtained
with the standard statistical approach.

\section{Ideal gas of spins}

This particular choice of the model, being extremely simple,
allows an analytical treatment of the case $N\to \infty $ in both
unperturbed and perturbed case. Let us analyze first the
unperturbed Hamiltonian $H_0$. This should be thought as a model
for a very weakly interacting system. In what follows, without loss
of generality, we assume $B=1$. The microcanonical ensemble
represents the most natural way to analyze an isolated system of
$N$ spins. One question is to find the thermodynamical temperature
in dependence on the energy $E\,$ . Another
is to study
the single particle energy distribution and the conditions under
which it can be assimilated to the Boltzmann distribution.

The temperature $T=\beta^{-1}$ ($k_B = 1$ hereafter)
can be defined via the
microcanonical ensemble:

\begin{equation}
\label{ent}\beta _{mc}={\frac{{\ dS}}{{dE}}}
\end{equation}
where $S$ is the entropy. For a sufficiently large number of
particles this can be in turn defined as $S=\log \rho (E)$.
Here $\rho (E)$ can be
defined through the phase space volume
\cite{H87}. If the motion of $N$
spins is ergodic, each of them covers uniformly the unit 3--d
sphere. Therefore, each component of ${\vec S\,}$has a uniform
probability density function in the interval $[-1,1]$\cite{I90},
that is~:

\begin{equation}
p(h) = {{ \partial P(S_i^z \leq h) }\over {\partial h}} =
\left\{
\begin{array}{ll}
1/2 \    {\rm when } -1\leq h\leq 1\\ 0 \  {\rm elsewhere} \\
\end{array}
\right.
\label{1p}
\end{equation}
and the same for the other components $S_i^y$ and $S_i^x$. Here
the quantity $P(x\leq a)$ gives the probability that the
continuous random variable $x$ gets values less than $a$. In the
same way, the density of states can be evaluated as a
probability~:

\begin{equation}
\label{clt}\rho _0(E_0,N)={\frac{{\partial P(\sum_{i=1}^NS_i^z\leq E_0)
}}{{
\partial E_0}}}
\end{equation}
The distribution of the sum of $N$ independent random variables
can be obtained using the central limit theorem (when $N\to \infty
$),

\begin{equation}
\label{rho0}\rho _0(E_0,N)\simeq
{\frac {1}{\sigma_0 \sqrt{2\pi}}} \exp
\left(-{\frac{E_0^2}{2\sigma_0^2}}\right)
\end{equation}
where
\begin{equation}
\label{sigma0}
\sigma_0^2 = \frac{N}{3}
\end{equation}
 As a result, from Eq.(\ref{ent}) one  gets an  unusual {\it
microcanonical relation} for the energy vs. inverse temperature
\cite{F64}:

\begin{equation}
\label{mmc}\beta _{mc}=-{\frac{{\ 3E_0}}{{N}}}
\end{equation}
In  principal, the energy $E_0\,$ ranges from $E_{\min }<0$
to $E_{\max }=-E_{\min}>0$, therefore, in this model negative
temperatures are possible too. One should note
that typical physical systems have density of states
increasing with  energy, thus giving positive temperature. For
this reason, in what follows, we consider only positive
temperatures which correspond to the left (part of the symmetric)
energy spectrum, $E_0<0$.

It is important to notice that the density of states can also be
obtained for finite system of particles without invoking central
limit theorems. It can be done by a direct integration over the
phase space volume,

\begin{equation}
\begin{array}{ll}
\rho_0 ( E_0 , N ) =
\int_{-\infty}^{+\infty} \  d S_1^z \ p (S_1^z) \ldots \
\int_{-\infty}^{+\infty} \
d S_N^z\  p(S_N^z)\times \\ ~~~~~~~~~~~~~~~~~\\
\times \delta ( \sum_{i=1}^N S_i^z - E_0 ) = \\
~~~~~~~\\ = \int_{-\infty}^{+\infty} \ {{1}\over{2\pi}} \ d
\lambda\ e^{-i\lambda E_0} [ \int_{-1}^{1} \ d x\ {{1}\over {2}}\
e^{i\lambda x } ]^N =\\ ~~~~~~~~~\\ = { {N}\over {2^N}}
\sum_{k=0}^{M} (-1)^k { {(N+E_0 -2k)^{N-1}}
\over { k! (N-k)! } }
\label{exaro}
\end{array}
\end{equation}
where $M$ is the integer part of $(E_0+N)/2$.

Let us now concentrate on the ``single-particle energy
distribution" ({\it SPE-distribution}) defined as~:

\begin{equation}
\label{n0def}n_0(h|E_0)=N{\frac{{\ \partial P(S_1^z\leq
h|\sum_{i=1}^NS_i^z=E_0)}}{{\partial h}}}
\end{equation}
where $P(x\leq a|y=b)$ is the conditional probability.

This quantity is the classical analog of the quantum {\it
distribution of occupation numbers}, which gives the
number of particles occupying
 a single-particle
level with energy $E_0$. Correspondingly, the above classical
distribution determines the probability that {\it any} of $N$
spins has the energy $E_0$. From this definition, it is clear that
the SPE-distribution (\ref{n0def}) is normalized to the total
number of particles, and it defines the mean energy of {\it all}
particles, or, the same, the mean total energy of the system. In
this way, one can treat an isolated model in the same way as the
model in  contact with a heat bath (see details in
Refs.\cite{BGI98},\cite{I99}).

It is well known that for $N\to \infty $ the single particle
energy should be distributed according to Boltzmann's law. Under
suitable conditions this holds true in this model, too. Indeed,
one gets

\begin{equation}
\begin{array}{ll}
n_0 (h|E_0) = N \ { {p(S_1^z=h,\sum_{i=1}^N S_i^z = E_0  )
}\over{\rho_0 (E_0,N)}}=\\ ~~~\\ = N \ { {p(S_1^z=h,\sum_{i=2}^N
S_i^z = E_0-h) }\over{\rho_0 (E_0,N)}}=\\ ~~\\ = N \ { {p(S_1^z=h)
p(\sum_{i=1}^{N-1} S_i^z = E_0)}\over{\rho_0 (E_0,N)}}=\\
~~~~~~~~\\ = N \ { {\rho_0 (h,1) \rho_0 (E_0-h,N-1)}\over{\rho_0
(E_0,N)}}
\end{array}
\label{bbb}
\end{equation}
where the last equality is due to the independence of $S_i^z$ .
The quantity $p(S_1^z=h,\sum_{i=1}^{N-1}S_i^z=E_0)$ defines the
joint probability density function.

Substituting the expression (\ref{rho0}) into the above relation,
it is easy to obtain (in the limit of a very large $N\gg 1$)~:

\begin{equation}
\begin{array}{ll}
n_{0} (h|E_0) \simeq
\exp \left(
-{  { 3E_0^2}\over{2 N (N-1)} } -{ { 3h^2}\over{2 (N-1)} } + { {
3E_0 h}\over{ N-1} }
\right) \\
\end{array}
\label{b1}
\end{equation}
where $E_0<0$.

\begin{figure}
\epsfxsize 7cm
\epsfbox{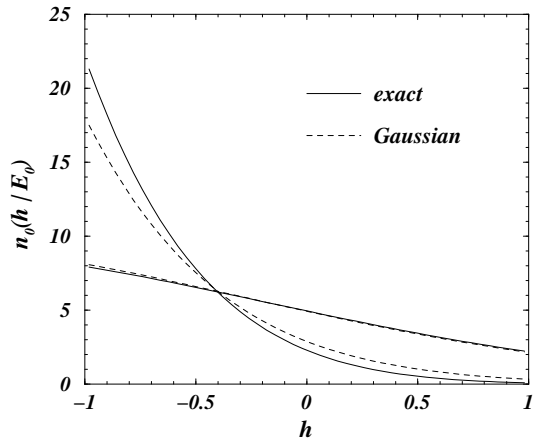}
\narrowtext
\caption{ Single particle energy distribution for $N=10$, and $J=0$.
Full lines represent the exact distribution (\protect\ref{bbb})
obtained from the exact unperturbed density of states
(Eq.(\protect\ref{exaro})). Dashed lines are the correspondent
infinite approximations (\protect\ref{b1}) with a proper
normalization. Upper curves are   for $E_0=-6$, lower ones for
$E_0=-2$. }
\label{n10e0}
\end{figure}

Obviously, a correspondent expression for any number of particles
can be obtained, using the exact value of the unperturbed density
of states Eq.(\ref {exaro}). In Fig.\ref{n10e0} we show the
infinite (Gaussian) approximation $(N\rightarrow \infty )$ and the
exact one, computed for two different energy values. As one can
see while in the middle of the spectrum the distributions are
almost the same, close to the band edges they are remarkably
different. The Gaussian, or infinite approximation, works very well
even for a small number of particles as soon as the energy $E_0$
is not close to the edges of the band (the interval $[-N,N]$ for
non-interacting spins). Indeed, the Gaussian density of states has
infinitely long tails, while the exact one is sharply defined
within $[-N,N]$. On the other hand while the support of the
density function scales as $N$, its variance depends on
$\sqrt{N}$. It follows that the region close to the edges becomes
less and less important as $N$ becomes large.

Let us briefly comment Eq.(\ref{b1}). Although it has been
obtained for a large number of particles, nevertheless it shows
that, generally speaking, the single particle energy distribution
for finite weak--interacting systems {\it is not necessarily
described by an exponential law}. One can see that the latter
occurs in a very strong limit $N\to \infty $, provided that $
|E_0|\gg |h|$ ,

\begin{equation}
\label{BD}n_0(h|E_0)\simeq \exp (-\beta _0h)
\end{equation}
where the temperature $T_0=\beta_0^{-1}$ is defined via
\begin{equation}
\label{beta0}
\beta_0 = -\frac{3E_0}{N-1} = \beta_{mc} + O(1/N)
\end{equation}
and $E_0<0$. One should
stress that, apart from the limit $N\rightarrow \infty $ , the
exponential distribution arises when the second condition $E_0\gg
|h|\sim 1$ is fulfilled. Physically this means that the total
energy must be larger than the typical single particle energy, a
condition naturally satisfied for a thermodynamical system.

Another important relation can be obtained from Eq.(\ref{bbb}).
Specifically, taking the derivatives of both sides of
Eq.(\ref{BD}) over $h$ , and exchanging the derivative in the
r.h.s., one gets~:

\begin{equation}
\label{bin}-{\frac{{d}}{{dh}}}\log n_0(h|E_0)={\frac{{d}}{{dE_0}}}\log \rho
_0(E_0-h,N-1)
\end{equation}
One can see that in the limit $N\to \infty $ and $|E_0|\gg |h|$,
the r.h.s coincides with the microcanonical definition of
temperature while the l.h.s. shows that the only possible
exponential distribution should have the microcanonical
temperature.

We have thus found that in the non-interacting system, when the
number of particles becomes sufficiently large and the total
energy of the system is larger than the typical single particle
energy, the {\it microcanonical temperature} (\ref{mmc}) defined
by the density of states, coincides with the {\it statistical
temperature} defined directly from the SPE-distribution
(\ref{n0def}) which is the standard Boltzmann law. One can then
assume that if the interaction is sufficiently small, in order not
to change dramatically the previous results, but enough to produce
ergodicity (from the equations of motion), Boltzmann's law again
results from taking a sufficiently large but finite number of
particles.

Let us stress that in order to get these results, the motion has
been assumed to be fully ergodic on the unit 3--d sphere.
Rigorously speaking, this can be true, from the dynamical point of
view, only in the presence of interaction $J\neq 0$. Indeed for
$J=0$ the motion is foliated onto regular tori and no ergodicity
at all is allowed. So, this ideal spin model has been worked out
following the traditional statistical mechanics picture, where the
interaction is assumed to be sufficiently weak in order to apply
non--interacting results, but sufficiently
strong in order to get ergodicity.

\section{Many interacting spins }

\subsection{Density of states}

\subsubsection{Infinitely strong interaction}

Before switching to the interacting case, let us compute the
density of states in the presence of an infinitely strong interaction
(namely putting formally $B=0$). In this case the density of
states can be written as~follows,

\begin{equation}
\label{rhoi}\rho _V(E,N)={\frac{{\partial P(V\leq E)}}{{\partial E}}}=
{\frac{
{P(J\sum_{i=1}^NS_i^yS_{i+1}^y\leq E)}}{{\partial E}}}
\end{equation}
The central limit theorem can also be applied in this case,
keeping in mind that the normalized
probability distribution  of
the product $z=xy$ of two uniformly distributed random variables
$x,y$ in the interval $(-1,1)$ is given by \cite{G74},
\begin{equation}
\label{1roi}\rho _V(E/J,1)=-(1/2J)\log (|E/J|)
\end{equation}
so that the variance is $\langle z^2\rangle =1/9$. Eq.(\ref{rhoi})
then becomes (in the large $N$-limit)~:

\begin{equation}
\label{rhoi2}\rho _V(E,N)\simeq
{\frac{1}{\sigma_V^2 \sqrt{2\pi}}} \exp \left(
-{\frac{E^2}{2\sigma_V^2}} \right)
\end{equation}
with
\begin{equation}
\label{sigmaV}
\sigma_V^2 = \frac{NJ^2}{9}
\end{equation}
 Note, that even in this case an explicit integral equation
can be obtained for finite systems. Following the same steps as in
Eq.(\ref{exaro}), one gets~:
\begin{equation}
\label{exaroi}\rho _V(E,N)=\int_{-\infty }^{+\infty }\ d\lambda \
e^{-i\lambda E}\ \left[ {\frac{{s(\lambda )}}{{\lambda }}}\right]
^N
\end{equation}
where $s(\lambda )$ is the sine-integral function defined by $$
s(\lambda )=\int_0^\lambda \ dx\ {\frac{{\sin x}}{{x}}} $$

\subsubsection{General case}

In the same way the density of states in the presence of both the
interaction and the ``kinetic'' term can be computed assuming
$H_0$ and $V$ are independent. Indeed, let us define~:

\begin{equation}
\label{rho}P(H_0+V\leq E)=\int_{-\infty }^E\ dE^{\prime }\ \rho (E^{\prime
})
\end{equation}
On the other hand, if $H_0$ and $V$ are independent, their joint
probability density function can be written as
\begin{equation}
\label{indep}\rho _{H_0,V}(E_0,E)=\rho _0(E_0)\rho _V(E)
\end{equation}
The probability density function of the sum of two independent
random variables is thus given by\cite{G74}~:

\begin{equation}
\label{rof}\rho (E)=\int_{-\infty }^{+\infty }\ dE^{\prime }\rho
_0(E-E^{\prime })\rho _V(E^{\prime })
\end{equation}
Substituting  Eqs.(\ref{rho0},\ref{rhoi}), and performing
simple Gaussian integrals one gets~:

\begin{equation}
\label{rho2}\rho (E,N)\simeq \frac{1}{\sigma \sqrt{2\pi}}
\exp \left( -{\frac {E^2}{2\sigma^2}} \right)
\end{equation}
where the variance
\begin{equation}
\label{sigma}
\sigma^2 = \frac{N}{3v} \,\,,\,\,\,\,\,\,
v=\frac{1}{1+J^2/3}\,\,,
\end{equation}
should be compared with Eq.(\ref{sigmaV}).

In Eq.(\ref{rho2}) there are two different approximations. The
first one is the Gaussian form for the density of states, which is
valid when $N\to \infty $. The second is the independence of the
(random) terms $H_0$ and $V$ which, of course, cannot be true in
general. For instance, a configuration with all spins aligned
along the $z$-axis, for which $E_0=N$ implies $V=0$. However, if
the motion is ergodic and the energy is not too close to the band
edges, the second assumption can be considered as a good
approximation.

\subsection{Classical analogs of quantum eigenstates and the LDOS}

Another important relation can be obtained linking the two
quantities, {\it shape of ``eigenfunctions}'' (SE) and {\it local
density of states} (LDOS), whose concepts were motivates by
quantum mechanics (see for example, \cite {FI97,WIC98,BGI98,I99}).
The latter quantity, LDOS, also known in nuclear physics as {\it
strength function}, is very important when describing the spread
of the energy initially concentrated in a specific unperturbed
state. The classical analogs of these functions have been
introduced in Ref.
\cite{CCGI96} and recently applied to dynamical models in
Refs.\cite{BGI98,WIC98}. The very point is that in the limit of
$N\to \infty $ these two quantities can be explicitly found and a
relation between them can be established.

In the case of ergodic motion the classical analog of the shape of
eigenstates (SE) can be defined as

\begin{equation}
\label{ef}W_E(E_0)={\frac{{\partial P(H_0\leq E_0|H_0+V=E)}}{{\partial E_0}}}
\end{equation}
Correspondingly, the classical analog of the LDOS is

\begin{equation}
\label{ldos}w_{E_0}(E)={\frac{{\partial P(H_0+V\leq E|H_0=E_0)}}
{{\partial E}}}
\end{equation}
It is very important that both the SE and LDOS can be computed
more efficiently \cite{CCGI96} using the equations of motion. For
the SE one has to choose a chaotic trajectory at some fixed energy
$E$, compute the $H_0(t)$ trajectory and sample the values of the
unperturbed Hamiltonian $H_0$ along this trajectory at some fixed
time intervals. This procedure gives us the ergodic distribution
inside the {\it energy shell} constructed by a projection of the
phase space of $H$ onto $H_0$ (see details in \cite{BGI98,WIC98}).
In the same way, the classical LDOS can be numerically computed taking
a bunch of (regular) trajectories of the unperturbed Hamiltonian
$H_0=E_0$ and computing the correspondent spread of $H(t)$ along
these unperturbed trajectories. The sample of the values of the total
Hamiltonian $H(t)$ taken at given intervals of time results in the
classical LDOS. Let us stress that even in the case of ergodic
motion of the total Hamiltonian $H$ , when only one single
trajectory is needed in order to get the SE, an ensemble of
trajectories of $H_0$ is necessary in order to get a reliable
result for the LDOS. This is a consequence of the integrability of
$H_0$.

Alternatively one can choose, as indicated by the definitions
(\ref{ef}),(\ref{ldos}), many different initial conditions on the
same energy surface and sum over them. It is clear that the two
procedures (dynamical and taking the average over the phase
volume) should give the same result in the case of ergodic motion.

An important identity can be easily proven from the relations
(\ref{ef}),(\ref{ldos}). Let us write,
\begin{equation}
\begin{array}{ll}
P(H_0+V \leq E , H_0 \leq E_0) =\\ ~~~\\ = \int_{-\infty}^{E_0} \
dE_0'
\ P(H_0+V \leq E, H_0=E_0') =\\
~~\\
\int_{-\infty}^{E_0} \ d E_0'\ P(H_0+V\leq E| H_0=E_0') \rho_0(E_0')
\end{array}
\label{id1}
\end{equation}
from which one can get the classical SE,

\begin{equation}
\begin{array}{ll}
w_{E_0} (E) \rho_0 (E_0) = { {\partial^2 P(H_0+V\leq E , H_0\leq
E_0)}
\over
{\partial E \partial E_0} }
\end{array}
\label{id2}
\end{equation}
In the same way one can obtain the classical LDOS,
\begin{equation}
\begin{array}{ll}
W_{E} (E_0) \rho (E) = { {\partial^2 P(H_0\leq E_0 , H_0+V \leq
E)}
\over
{\partial E_0 \partial E} }
\end{array}
\label{id3}
\end{equation}

From above, the following relation emerges,

\begin{equation}
\label{id}w_{E_0}(E)\rho _0(E_0)=W_E(E_0)\rho (E)
\end{equation}

Let us stress that the previous identity does not depends on
the Gaussian approximation and it takes into account dynamical
correlations too.
This simple relation between classical SE and LDOS is very
important in different applications. Remarkably, only unperturbed
and perturbed densities of states, $\rho_0(E)$ and $\rho(E)$, get
into this relation. This allows to relate the shape of eigenstates
to that of the LDOS in the corresponding quantum model, in a deep
semiclassical region. In fact, the knowledge of these two
functions leads to a novel semiclassical approach according to
which it is easy to detect quantum effects of localization, see
details in \cite{FI97,I99}.

It should be stressed that classical SE and LDOS, in essence, are
ergodic measures for energy shells defined by the projection of $H_0$
onto $H$ (and vice versa) in the energy representation.

Functions $W_E(E_0)$ and $w_{E_0}(E)$ can be considered as kernel
operators transforming unperturbed quantities in total ones and
vice versa. For instance, by integrating Eq.(\ref{id}) one has~:
\begin{equation}
\label{p1}\rho (E)=\int \ dE_0\ w_{E_0}(E)\rho _0(E_0)
\end{equation}
and the converse,

\begin{equation}
\label{p2}\rho _0(E_0)=\int \ dE\ W_E(E_0)\ \rho (E)
\end{equation}
In a certain way $w$ and $W$ can be considered one the inverse of
the other. It is also easy to check that when $J\to 0$ then
$W_E(E_0)=w_{E_0}(E))=\delta (E-E_0)$.

Assuming the Gaussian approximation for the densities of states,
and using Eq.(\ref{id}) we can easily get an analytical expression
for the classical SE~:

\begin{equation}
W_{E} (E_0 ) = \frac{1}{\sigma_W \sqrt{2\pi}}
\ \exp \left[ -{{(E_0 - E_c )^2}\over{2\sigma_W^2}} \right]
\label{ef1}
\end{equation}
where
\begin{equation}
\label{sigmaW}
\sigma_W^2 = \frac{NJ^2}{9} v
\end{equation}
is the variance, and $E_c=vE$ is the center of the SE.

In order to obtain the LDOS distribution, one can use the relation
(\ref{id}) from which one gets,
\begin{equation}
\label{ldosgauss}
w_{E_0} (E ) = \frac{1}{\sigma_w \sqrt{2\pi}}
\ \exp \left[ -{{(E_0 - E )^2}\over{2\sigma_w^2}} \right]
\end{equation}
where
\begin{equation}
\label{sigmaw}
\sigma_w^2 = \frac {NJ^2}{9}
\end{equation}
It is interesting to note that the above distribution, in
the Gaussian approximation, coincides
with Eq.(\ref{rhoi2}),

\begin{equation}
\label{roi2}w_{E_0}(E)=\rho _V(E-E_0).
\end{equation}
This relation can be also obtained independently, using the
assumption of the independence of $H_0$ from $V$.

 One can see that although
the two Gaussians are different, for a weak interaction $J\ll 1$,
 and then $v\sim1$,
they appear to be close one to each other. This fact is of a
general nature and occurs in other models, see
Refs.\cite{CCGI96,WIC98,BGIC98}.

On the other hand, this result shows that, strictly speaking,
even in the case of ergodic motion, one should not expect SE and
LDOS to be the same. As was found above, the relation between the
SE and LDOS is given by Eq.(\ref{id}). One can obtain a very useful
relation for the variances of the SE and LDOS (valid in the Gaussian
approximation only),
\begin{equation}
\label{relation}
\frac{\sigma_w^2}{\sigma_W^2} = \frac{\sigma_0^2}{\sigma^2}.
\end{equation}

We would like to stress again that both  SE and LDOS have a proper
meaning only in the case of ergodic motion. In other cases they
strongly depend on initial conditions and on the energy, thus not
allowing to use any statistical approach.

\subsection{Distribution of single particle energies and different
temperatures}

For $N\gg 1 $, the microcanonical temperature $T_{mc}=\beta ^{-1}$
can be defined from the total density of states (\ref{rho2}),

\begin{equation}
\label{mci}\beta _{mc}={\frac{{\partial \log (\rho )}}{{\partial E}}}=
-{\ \frac{3E}{N}} v
\end{equation}
with $v=1/(1+J^2/3)$. In fact, Eq.(\ref{mci}) determines the {\it
thermodynamical temperature} since it establishes the relation
between the temperature and total energy of a system. Now, we are
going to find the {\it statistical temperature} associated with
the SPE-distribution. First of all, let us notice that, if the
variables $H_0$ and $V$ are independent the following approximate
relation can be obtained~:

\begin{equation}
\begin{array}{ll}
n(h|E) = { {dP(S_1^z\leq h | H_0+V=E)}\over{dh}}=\\ ~~\\ = \int \
dE_0 \ n_0(h|E_0) \  W_E(E_0)
\end{array}
\label{pir}
\end{equation}
Substituting Eqs.(\ref{b1},\ref{ef1}), in Eq.(\ref{pir}) one gets,

\begin{equation}
\label{bint}
\begin{array}{ll}
n(h|E)\simeq \exp \left( -{\frac{{\ 3E_c^2}}{{
2N(N-v)}}}-{\frac{{\ 3vh^2}}{{2(N-v)}}}+{\frac{{\ 3E_ch}}{{\
N-v}}}\right) &
\end{array}
\end{equation}
where  $E_c=vE$. Note that when $J=0$, then $v=1$ and Eq.(
\ref{b1}) is recovered.

One can see that in the limit $|E|\gg |h|\sim 1$ and  $N\gg v$ the
SPE-distribution can be approximated by the exponential
dependence:~
\begin{equation}
\label{ff}n(h|E)\sim e^{-\beta (E)h}
\end{equation}
where
\begin{equation}
\label{equa}\beta (E)= - \frac {3E}{N}v=\beta _{mc}
\end{equation}

The last equality, claiming that the statistical temperature
coincides with the thermodynamical one, is by no means trivial.
From one side, the presence of strong interaction suggests some
statistical equilibrium property thus leading to the
microcanonical predictions. On the other side we do not know if
the single particle description is valid since the interaction is
strong. Even in the case when mean field approach would be
possible (in such a way that part of the interaction becomes
actually a single particle energy), the single-particle
distribution would depend on the mean field and there are no
reasons to expect, a priori, that
the Boltzmann distribution
occurs with the same temperature as given by the microcanonical
ensemble.

Another way to find the statistical temperature which corresponds
to the SPE-distribution, is as follows. First, we note that in the
presence of interaction one has

\begin{equation}
\label{ec}\int \ dh\ h\ n(h|E)=\bar E\ne E
\end{equation}
that is different from the non-interacting case, for the latter
the obvious relation holds,

\begin{equation}
\label{ec0} \int \ dh \ h \ n_0 (h | E_0 ) = E_0
\end{equation}

Note that the SPE-distribution in Eq.(\ref{ec}) depends on the
interaction, however, the ``mean value" of the energy of all
particles does not correspond to the energy of a system. This very
fact allows to relate
statistical effects of the
interaction  to an increase of the total energy
(in application to quantum systems this approach was
considered in \cite{FI97,FI97a,I99}).

Specifically, in order to find the SPE-distribution $n(h|E)$ from
Eq.(\ref{ec}), one needs to know the {\it renormalized energy}
$\bar E$. One can see that the Boltzmann distribution gives a
correct result for an isolated system, if we take into account the
shift of the energy $\Delta_E = \bar E - E$ which is due to the
interaction between particles. In a sense, the (statistical) effects
of the interaction are absorbed by the increase of the energy of a
system, compared to the case of non-interacting particles.
Therefore, pseudorandom interaction may be treated as an internal
heat bath, thus giving rise to a statistical equilibrium.

The shift $\Delta_E$ can be found by assuming the absence of
correlations between $H_0$ and $V$ in a way described above. One
can show that

\begin{equation}
\label{ecc}\bar E=\int \ dE_0\ E_0\ W_E(E_0)=E_c
\end{equation}
where $E_c=vE$ is the center of the classical SE, and the last
equality occurs in a strong limit $N \gg 1$ using Eq.(\ref{ef1}).
Therefore, the shift is given by the following relation,
\begin{equation}
\label{shift}
\Delta_E = \bar E - E = -\frac{\sigma_w^2}{\sigma_0^2} E = -
\frac{J^2}{3} E
\end{equation}

It is important that the shift is defined by the width $\sigma_w$
of the LDOS
and by the width of the unperturbed density. One should note that in quantum
representation, the variance $\sigma_w^2$ is defined by the sum of
squared off-diagonal matrix elements. Therefore, this
energy shift can be found without diagonalization of huge quantum
Hamiltonian matrices. Using this shift, one can find the
temperature from Eq.(\ref{ec}) by assuming the Boltzmann
dependence for $n(h)$.

One more way to obtain the temperature is via the unperturbed
density of states evaluated at the renormalized energy $\bar E$

\begin{equation}
\label{rene}\beta _r=\left. {\frac{{d\log \rho _0}}{{dE_0}}}\right| _{E_0={%
\bar E}}=-{\frac{{3\bar E}}{{N}}}
\end{equation}
Using the relation between $\bar E$ and $E$, one gets,
\begin{equation}
\label{rene1} \beta_{r} = - {\frac{{3 \bar{E} }}{{N}}} =
 - {\frac{3E}{N}}v = \beta_{mc}
\end{equation}

Once again, we should remind that apart from the limit $N \gg 1$,
the above relation is valid when neglecting the correlations
between $H_0$ and $V$. However, the smaller is the number of
particles, the larger is the energy region where (dynamical)
correlations will be important.

It is now interesting to find the increase of temperature $\Delta
T=T-T_0$ due to the interaction, in comparison with the
temperature $T_0=\beta_0^{-1}$ of the system with non-interacting
spins,
\begin{equation}
\label{deltaT}
\frac{\Delta T}{T_0} = \frac {\sigma_w^2}{\sigma_0^2} = \frac
{J^2}{3}
\end{equation}
One can see that the relative increase of temperature is given in
terms of the variance of the LDOS and the width of the unperturbed
density only (see also Eq.(\ref{shift}).
This fact seems to be generic, see discussion in
\cite{FI97,I99}.

 More accurate relation
for the definition of temperature via the SPE-distribution,
without the assumption of the Gaussian form of the density of
states, can be obtained as follows. In close analogy to
Eq.(\ref{bin}) one can write,
\begin{equation}
\begin{array}{ll}
n(\epsilon|E) = { {\partial P(S_1^z\leq \epsilon| H=E)}
\over{\partial \epsilon}}=\\
~~\\ ={ {1}\over{\rho(E,N)}} { {\partial P(S_1^z\leq \epsilon,
H=E)}
\over{\partial \epsilon}}
\end{array}
\label{z1}
\end{equation}
On the other hand, defining $x_i=S_i^z$, and $y_i=JS_i^yS_{i+1}^y$
it follows~:
\begin{equation}
\begin{array}{ll}
P(x_1\leq \epsilon, H=E)=\\ ~~~~~~~\\ =\int_{-\infty}^{\epsilon} \
d\epsilon'  P(x_1 =\epsilon', x_1 + y_1 + \sum_{i=2}^N x_i + y_i
=E)=\\ ~~~~~~~\\ =\int_{-\infty}^{\epsilon} \ d\epsilon'  P(x_1
=\epsilon', y_1 + \sum_{i=2}^N x_i + y_i =E-\epsilon')=\\ ~~~~~~\\
= \int dh
\int^{\epsilon} d\epsilon' P(x_1 =\epsilon',
y_1=h,  \sum_{i=2}^N x_i+y_i=E-\epsilon'-h)
\end{array}
\label{z2}
\end{equation}

Owing the independence of the 3 random variables, the joint
probability density can be factorized and one gets~:
\begin{equation}
\begin{array}{ll}
n(\epsilon|E)\rho(E,N)=
\\
~~~~~~~~\\ = \rho_0(\epsilon,1) \int \ d h \
\rho_V(h,1) \rho(E-\epsilon-h,N-1)
\end{array}
\label{z3}
\end{equation}
Defining now the ``interacting'' density of states

\begin{equation}
\begin{array}{ll}
Z_{\epsilon}(E) =
\int \ d h \
\rho_V(h,1) \rho(E-\epsilon-h,N-1)
\end{array}
\label{z4}
\end{equation}
one gets a relation similar to Eq.(\ref{bin})~:
\begin{equation}
\label{z5}
-{\frac{{\ \partial }}{{\partial \epsilon }}}\log n(\epsilon |E)={\
\frac{{\ \partial }}{{\partial E}}}\log Z_\epsilon (E)
\end{equation}

Eq.(\ref{z5}) states that if correlations can be neglected then a
renormalized density of states must be introduced in order to get
the ``correct"  temperature  as obtained from the
single particle energy distribution. On the other hand, it is
clear that in the limit $N\to
\infty $, and $|E|\gg 1$, we have $Z_\epsilon (E)\sim \rho (E)$ and the
usual definition is recovered. It is now interesting  to apply
our estimates to a system with not very large number of spins.

\begin{figure}
\epsfxsize 7cm
\epsfbox{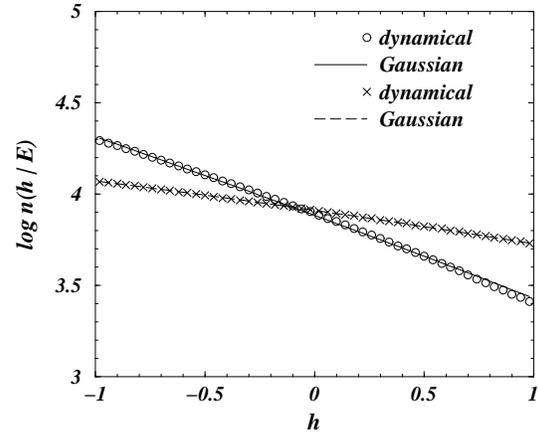}
\narrowtext
\caption{  Single particle energy distributions
for $N=100$ and large interaction $J=1$, for different energy
values. One single trajectory has been iterated for a given
energy, up to a time $t= 10^6$. Numerical data should be compared
with the analytical expression (\protect\ref{bint}) obtained from
the Gaussian approximation for the density of states. Circles
stand for $E=-19$, crosses for $E=-7$.}
\label{n100e}
\end{figure}

\section{Numerical data}

\subsection{Large number of spins}

Let us first consider the model with $N=100$ spins. On one hand,
this situation is far from the thermodynamic limit, on the other
hand, the number of particles is quite large and one can expect a
good correspondence with analytical results obtained in previous
Section. In Fig.\ref{n100e} two SPE-distributions are plotted for
different values of the total energy $E$. In both cases, the
trajectories have been found to be chaotic with positive maximal
Lyapunov exponents. For comparison, the analytical expression (see
Eq.\ref{bint}), obtained in the Gaussian approximation, is shown.
As one can see, the agreement is fairly good.

\begin{figure}
\epsfxsize 8cm
\epsfbox{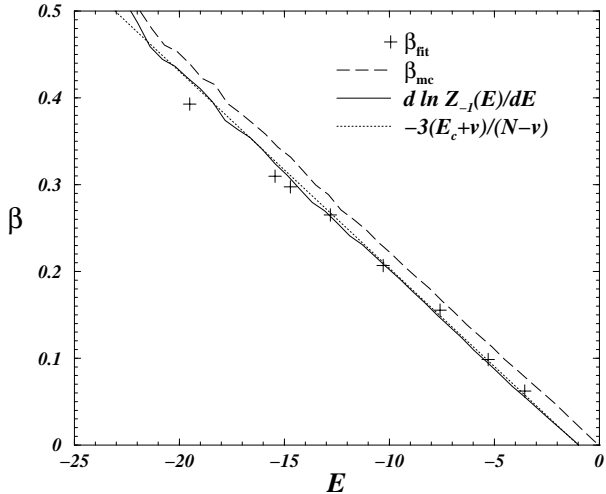}
\narrowtext
\caption{  Inverse temperatures versus total energy for $N=100$ and $J=1$.
Crosses are the extracted $\beta_{fit}$ from the best fit of the
SPE-distributions $n(h | E)$ to the exponential dependence.
Typical examples of these distributions are given in
Fig.\protect\ref{n100e}. Dashed line is the microcanonical
relation (\ref{mci}) found numerically from the density of states
$\rho(E)$. Full line is the theoretical prediction, obtained from
Eq.(\protect\ref{z5}) with $\epsilon=-1$. Dotted line is the
Gaussian approximation (see in the text).}
\label{n100b}
\end{figure}

On the other hand, a direct comparison of the microcanonical
temperature $\beta_{mc}$ with the approximated temperature,
obtained by fitting the SPE-distribution by an exponent, gives
different answers, see Fig.\ref{n100b}.

 In this figure crosses
represent the fitted inverse temperatures as a function of the
total energy, while the dashed curve is the microcanonical
relation between the inverse temperature and the energy. In order
to smooth fluctuations, the derivative has been calculated
performing local averages in small energy windows. A small,
but systematic,
difference between the statistical and microcanonical temperatures
is clearly seen.

A much better agreement can be achieved by using directly,
(instead of the total density of states) the renormalized density
Eq.(\ref{z4}). The
corresponding result is presented by the full line. One can see
quite good correspondence to numerical data for the statistical
temperature. This result means that  Eq.(\ref{z4}) is
more accurate than the usual statistical definition
(\ref{mci}).

One can also introduce the effective (statistical) temperature by
making use of Eq.(\ref{bint}), keeping all terms in the
exponent. The temperature can be defined as a slope of $\ln
n(h|E)$ at the bottom of the energy spectrum,
\begin{equation}
\label{lnslope}
\beta_{fit} = \frac {d \ln n(h|E)}{d h} = \frac{3E_c}{N-v}
-\frac{3vh}{N-v} = \frac{3v(E-h)}{N-v}
\end{equation}
with $h\rightarrow -1$. The corresponding result is shown in
Fig.\ref{n100b} by the dotted line. One can see that this
dependence practically coincides both with the fit to actual
distribution $n(h|E)$ (crosses), and with the temperature
determined by Eq.(\ref{z4}) (full line). This means that these two
approximations correctly take into account both the dynamical
correlations and the finite number of particles.

We can conclude that, in spite of the relatively  small number of
particles, if compared with the usual thermodynamical systems of
$10^{23}$ particles, our model of $N=100$ interacting spins can be
approximately described by standard statistical approach which
ignores dynamical correlations between particles. Indeed, the
difference between microcanonical temperature and the approximate
temperature found by the fit of the SPE-distribution close to the
edge of the energy spectrum, is quite small and may neglected in
some cases. However, even for a relatively large number of spins,
$N=100 \gg 1$, a clear influence of dynamical correlations and
finite number of particles remains.

One should, however, stress that in spite of a clear manifestation
of the influence of dynamical correlations and finite number of
particles, the SPE-distribution $n(h|E)$ for low energies can be
effectively described by the standard Boltzmann distribution,
however, with a renormalized temperature.

\subsection{Chaos versus ergodicity}

It is reasonable to think that a statistically stable distribution
would require a certain degree of chaoticity. But chaos itself, as
indicated, for instance, by the positivity of the maximal Lyapunov
exponent, is not enough in order to get equipartition among
different degrees of freedom (see general discussion of this very
important problem in
\cite{Z99}).
 We have found such cases for a small number of particles and  energy
close to the center of the spectrum.

\begin{figure}
\epsfxsize 8cm
\epsfbox{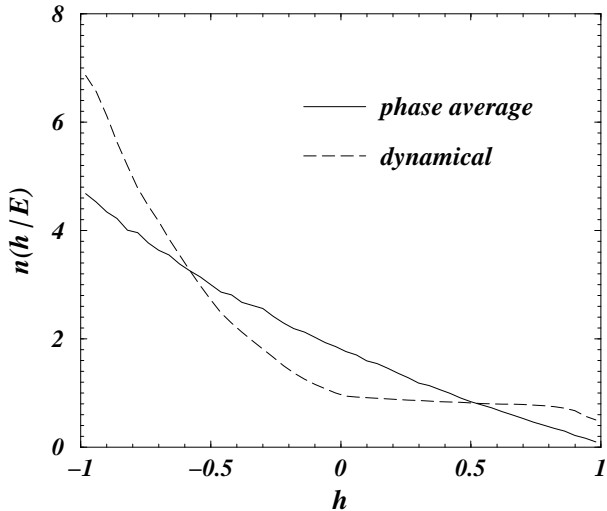}
\narrowtext
\caption{ Single particle energy distribution for $N=4$, and $J=1$.
Dashed line is obtained dynamically, by integrating the equations
of motion for a single trajectory with energy $E_*=-1.9745$ up to
the time $t=10^7$. Full  line is the phase average distribution
obtained for $10^6$ different initial conditions in the energy
range $[-1.98,-1.97]$. }
\label{n4}
\end{figure}

It is instructive to work out a specific example. Let us consider
a system of $N=4$ spins. For $J=B=1$ and energy $E_*=-1.9745$  the
single particle energy distribution is very different from the
phase average distribution (obtained with many different initial
conditions in a small energy window close to $E_*$), see
Fig.\ref{n4}. Nevertheless, the Lyapunov exponent is positive for
many initial conditions inside the same energy window.

This lack of ergodicity is also reflected in the
lack of equipartition. Indeed, defining the
average unperturbed single particle  energy, as
\begin{equation}
\langle S_i^z \rangle = \lim_{T\to\infty} \int_0^T \ dt \ S_i^z(t)
\label{ke}
\end{equation}
we get that, typically, for trajectories with energy $E$ close to
$E_*$,  neighbors spins do not share the same energy (lack of
equipartition).
 In Table I , we show the average kinetic energy per spin
for a dynamical trajectory with the energy $E_*$ (second column),
to be compared with the average kinetic energy per spin as
obtained from the phase average distribution (right column) within
the energy range $\Delta E = [-1.98,-1.97]$.

\begin{table}
\caption{Average kinetic energy per spin.
}
\begin{tabular}{lrl}
 spin label $i$
&  $\langle S_i^z(E_*) \rangle $
&  $\langle S_i^z(\Delta E) \rangle   $\\
\tableline
$1$&$-0.7248    $&$  -0.3783   $\\
$2$&$-0.1387    $&$  -0.3788   $\\
$3$&$-0.7248    $&$  -0.3788  $\\
$4$&$-0.1388    $&$  -0.3801  $\\
\end{tabular}
\label{tabl}
\end{table}
\vspace{0cm}

\subsection{Ergodicity versus dynamical correlations}

In this section we deal with systems having a small number of
spins (let us say, order of ten). In contrast to the case
considered in the previous Section, we select only those energy
values which correspond to both a positive Lyapunov exponent and
to equipartition among different spins; this means, in
particular, that  ``dynamical'' and ``statistical'' distributions
(the former obtained by integration of equations of motion for one
trajectory, the latter by choosing many different initial
conditions onto the energy surface and performing a phase average)
are close one to each other.

\begin{figure}
\epsfxsize 7cm
\epsfbox{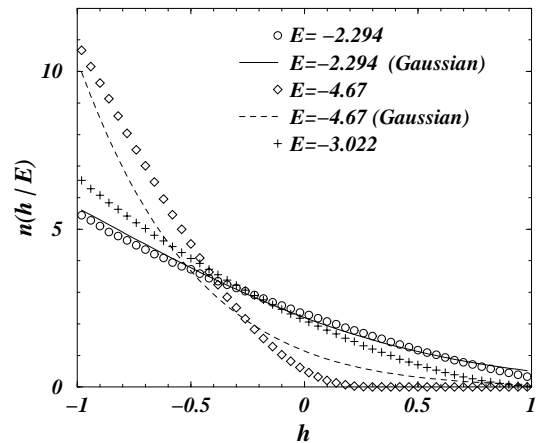}
\narrowtext
\caption{ Single particle energy distribution
$n(h|E)$
for $N=5$
with interaction  $J=1$ as a function
of single particle energy $h$.
Different symbols indicate
different energy values $E$, as shown in the window.
For few sets of symbols the Gaussian approximation, as given
by Eq.(\ref{bint}) has also been shown.
}
\label{n105}
\end{figure}

\begin{figure}
\epsfxsize 7cm
\epsfbox{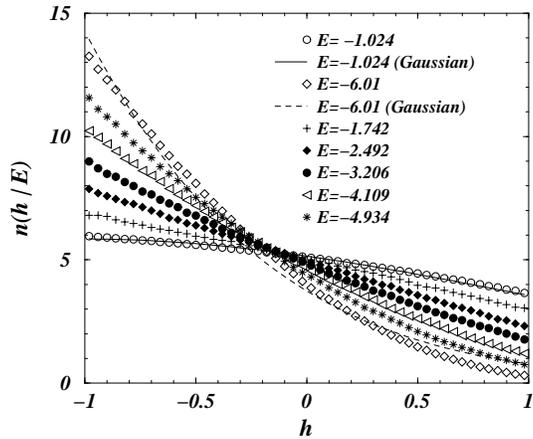}
\narrowtext
\caption{ The same
for $N=10$. }
\label{n106}
\end{figure}

Let us focus on two different samples of $N=5$ and $N=10$
interacting spins with strong and chaotic interaction ($J=1$). In
Figs.\ref{n105},\ref{n106} we show the SPE-distribution for
different energies, together with the correspondent Gaussian
approximations Eq.(\ref{bint}).

 The first important point is a remarkable deviation from
the Gaussian approximation, at least for energy values $E$ close
to the band edge. This simply means that the approximations
involved in order to get Eq.(\ref{bint}) are no longer valid. One
of these approximations was to consider a Gaussian shape for the
SE in Eq. (\ref{pir}). According to additional data,  the
effective SE (statistical or dynamical, they are very close one to
each other) for the same energy value show remarkable deviations
from the Gaussian. Nonetheless, the substitution of  the "true" SE
in Eq.(\ref{pir}) does not affect considerably  $n(h|E)$. This
amount to say that Eq.(\ref{pir}) itself does not constitute a
good approximation for energy values close to the edges.

Indeed, due to a small number of spins, and to the relatively
large energy shared by each spin particle, they are strongly
correlated and most of the uncorrelation assumptions (between
$H_0$ and $V$, for example) made in the previous chapters are no
longer valid.

To be more precise, it can be shown that Eq.(\ref{pir})
can be considered as the ``diagonal'' approximation of
the following exact relation

\begin{equation}
n(h|E) = \int \ dE_0 W_E (E_0) \ n_E (h|E_0)
\label{pire}
\end{equation}

where
\begin{equation}
 n_E (h|E_0) = P(S_1^z = h | H_0=E_0, H=E)
\label{coke}
\end{equation}
In the limit when $H$ and $H_0$ are independent one recovers
Eq.(\ref{pir}) since $ n_E (h|E_0) = n_0 (h|E_0)$. The study of
the correlation kernel Eq.(\ref{coke}) beyond the diagonal
approximation will be deserved for future investigations.

On the other side, in the middle of the energy band $|E| \sim 0$
there is a rough agreement with the Gaussian approximation (see
Fig.\ref{n105}). Indeed, for such energy values, there is no
preferred direction of the spin (the energy shared by each single
spin is relatively small) and Eq.(\ref{pir}) still represents a
good approximation. However, the bad point is that close to the
center of the band the Gaussian approximation for $n(h|E)$ is far
from an exponential (in fact, it is close to a Gaussian!, see
Eq.\ref{bint}).
Indeed, let us remember that one of the conditions in order to
obtain the Boltzman distribution was $|E|\gg1$, which is of course
not satisfied at the center of the energy spectrum.

Even if not completely satisfactory from the theoretical point of
view, but in close analogy with  the
case of $N=100$ spins, one could define a temperature as (minus)
the slope of the fitting straight line  to $\log n(h|E)$.
Operatively we observe that in the region to the left of their
intersection point ($h\sim -0.6$ for $N=5$ and $h\sim -0.7$ for
$N=10$), distributions with different energy value have a behavior
closer to an exponential. Therefore, it is natural to fit the
numerical $n(h|E)$ with an exponential only to the left of the
intersection point. Let us call $\beta_{fit}$ the inverse
temperature obtained in this way.

\begin{figure}
\epsfxsize 7cm
\epsfbox{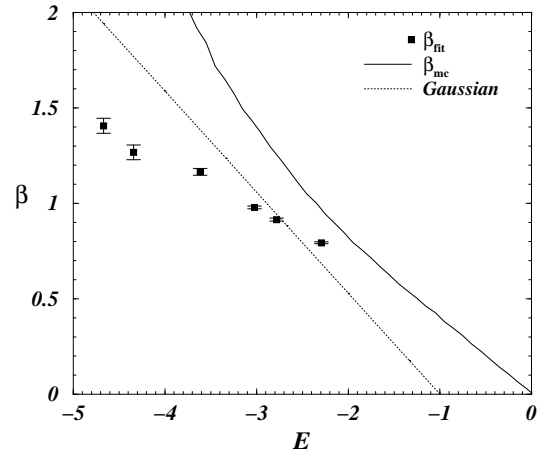}
\narrowtext
\caption{ Comparison between different definitions of temperature
for $N=5$. Symbols represent the numerically extracted
$\beta_{fit}$, while different curves refer to different
definitions of temperature,
microcanonical (full line) and the Gaussian approximation (dotted
line). Since the latter turns out to be dependent on $h$ too, we
computed it at $h=-1$. }
\label{n105f}
\end{figure}

\begin{figure}
\epsfxsize 7cm
\epsfbox{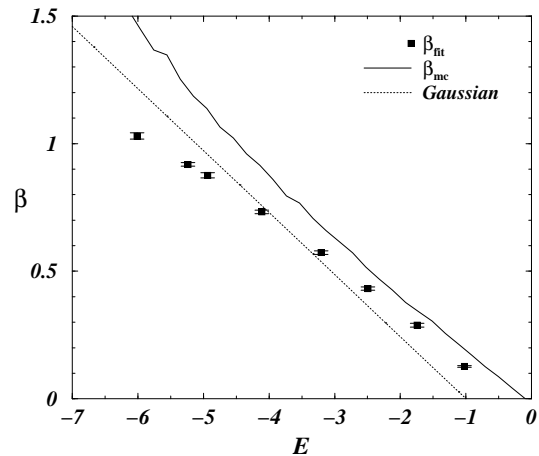}
\narrowtext
\caption{ The same for $N=10$ spins.
}
\label{n105g}
\end{figure}

Needless to say, when compared with both the statistical and
microcanonical temperature, one can see important deviations. In
Figs.\ref{n105f},\ref{n105g} we plot the obtained $\beta_{fit}$ as
a function of the energy $E$, together with other definitions of
temperature for $N=5$  and $N=10$.

As one can see, none of the previous definitions seems to fit the
numerical values. This is not surprising, in spite of the fact
that only an approximate exponential behavior has been found. We do
not have any approximate theory able to
describe such temperature differences when the number of spins is
small. Strictly speaking, one recognizes the importance of the
classical SE in the description of the behavior of single particle
distribution, but it becomes technically complicated to go beyond
the diagonal approximation, which is correct only when the number
of particles is sufficiently large and the energy is not too close
to the ``many--body ground state'' (bottom of the energy
spectrum).

\begin{figure}
\epsfxsize 7cm
\epsfbox{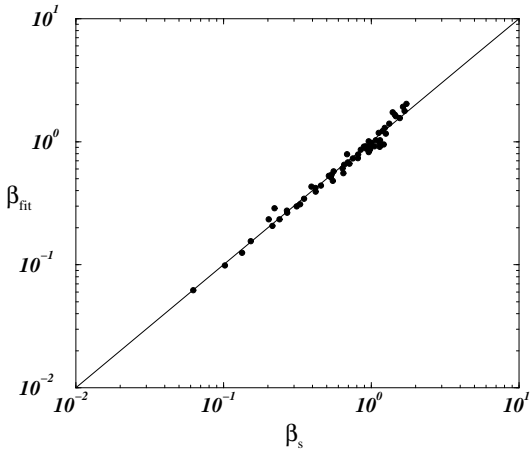}
\narrowtext
\caption{ $\beta_{fit}$ obtained by fitting the
single particle energy distribution
in the negative part of the single particle spectrum, versus
the proposed phenomenological scaling $\beta_s$. Full circles represent
the numerical data for $N=4,5,10,20,100$, and $J=0.1,0.5,1$ and
different energy. Full line is the  scaling relation.
}
\label{fscal}
\end{figure}

The following approximate phenomenological scaling relation has
been numerically found:

 \begin{equation}
\beta_{s} = -{{3  \langle E\rangle  }\over {N}} -
{{ (1+\alpha) J^2}\over{N}}
\label{scal}
\end{equation}
where $\alpha=0.7$ is the fitting parameter and $\langle E
\rangle =
\int \ dh \ h \ n(h|E)$ . In Fig.\ref{fscal} for
different $N$, $J$ and energy $E$ we show the numerical data and
the scaling relation Eq.(\ref{scal}).

We have not, for the time being, a theoretical explanation of this
scaling relation. It should be stressed that, in the limit
$N\to\infty$ and sufficiently large energy $\langle E \rangle$ the
second term in the l.h.s. of Eq.(\ref{scal}) is negligible with
respect to the first one and
Eq.(\ref{rene1}) is recovered. We remind that it is not possible
to take $J\gg1$ in this model since it becomes integrable, and
most of the previous results are necessarily wrong.

The disagreement between the microcanonical temperatures and the
statistical one for a small number of particles has few important
theoretical implications. First of all we note that they are
obtained in two completely different ways. While the single
particle distribution reflects a property of the constant energy
surface $H=E$, the microcanonical definition requires a derivative
{\it across} the energy surface. In principle, the knowledge of
the correlation kernel Eq.(\ref{coke}) would solve completely the
problem. But this in turn requires the knowledge of the (infinite)
intersections among the $H=E$ surface and the $H_0=E_0$ ones.

\section{Concluding remarks}

In this paper we have studied the emergence of  Boltzmann's law
for the single particle energy distribution $n(h|E)$ in a
isolated dynamical model of finite number of interacting spins. In
the limit of a very large number of spins this model allows an
analytical treatment. We have shown that in this strong limit,
Boltzmann's distribution, indeed, occurs with an effective {\it
statistical} temperature which coincides with that defined by the
standard {\it microcanonical temperature}. The latter is defined
via the total density of states.
Since our analytical proof is also valid for a strong
spin interaction, it is far from trivial. Indeed,
it is not { a priori } clear that the
SPE distribution, which pertains a non--interacting  property,
follows the Boltzmann
distribution with the temperature determined via the total density
of states.

The above result has been obtained in an approach which is very
similar to that recently suggested in the study of the so-called
{\it two-body random interaction model}\cite{FIC96,FI97,FI97a}.
 According to this approach, the distribution
of occupation numbers for single-particles levels in a quantum
many-body system, is directly related to the average shape of
chaotic eigenstates in the basis of the unperturbed Hamiltonian
(the latter can be considered as the mean-field part of a system).
This means that, in fact, there is  no needs to know exact eigenstates
of  huge Hamiltonian matrices which takes into account two-body
interaction between particles. This is due to the chaotic nature
of eigenstates, which results from the (assumed) randomness of the
two-body matrix elements.

An important point of the above approach is that in some cases the
average shape of eigenstates can be found analytically from
off-diagonal matrix elements of the total Hamiltonian $H_0$, see
Ref.\cite{FI97,FI99}. The same happens to another important
quantity, the {\it strength function}, which  in solid state
physics is also known as the {\it local density of states} (LDOS).
These two quantities are related to each other, however, so far
this relation is not well-studied. Numerical data for different
models, disordered
\cite{FI97} and dynamical \cite{WIC98,BGI98} as well, have shown that for
not very strong interaction, these two quantities are very close
one to each other. The knowledge of  LDOS is very important in
many applications. Indeed it gives the information on how energy, initially
concentrated in a specific unperturbed state, spreads over all
other states due to the interaction between particles. The inverse
width of LDOS is, in fact, the effective time of this spread.

Until recently, the above two quantities have been discussed only
in the context of quantum systems. On the other hand, in
\cite{CCGI96} it was noted that both  LDOS and SE have a very
clear classical analog. The study of the {\it classical} LDOS and SE
have been started in \cite{WIC98,BGI98}, and the first results
have shown that the average shape of quantum eigenstates
(and the same for the LDOS) for chaotic eigenstates  coincides
fairly well
with the classical counterpart. For this reason, when studying
the occurrence of  Boltzmann's distribution in our model of
interacting spins, we have also paid attention to the classical
LDOS and SE.

Our results in what concerns classical SE and LDOS have shown a
nice correspondence to the main findings for the quantum model
with random two-body interaction \cite{FI97,FI97a,I99}. In
particular, in our classical model we have proved the basic
expressions for an increase of both  total energy (\ref{shift})
and  temperature (\ref{deltaT}) due to statistical effects,
which in quantum model have been derived directly from the shape
of exact eigenstates in the unperturbed energy basis. Moreover,
new relations (\ref{id}) and (\ref{relation}) have been found
which may be important also in quantum systems.

Numerical data for our model have confirmed main theoretical
predictions for a large number of particles, $N=100$. However, it
was also found that in spite of a relatively large $N$, one can
detect clear deviations for the statistical temperature related
to the SPE-distribution. Detailed analysis shows that these
deviations are mainly due to dynamical effects of correlations
originating from a finite number of particles. As was shown,
analytical results can be interpreted as a ``diagonal''
approximation, that is neglecting correlations between the total
and the unperturbed Hamiltonian. One should stress that the
correction to the thermodynamical expressions are not only on the order
of $1/N$ as commonly assumed in the  literature, but depend on the
interaction strength as well.

Drastic deviations from the thermodynamical approach have been
found for few interacting spins ($N=10$ and $N=5$). Namely, both the
thermodynamical temperature defined from the microcanonical
relations, and the statistical temperature found analytically for
a very large $N$, are very different from the approximate
temperature. The latter has been determined numerically from the
Boltzmann dependence of the SPE distribution, used as a fitting
expression for low energies of the system. A phenomenological
expression has been found from the analysis of the data, however,
without a proper analytical explanation. These problems, as well
as the developments of a {\it semiquantal} approach for which
quantum distribution of occupation numbers is computed with the
use of classical analog of the shape of eigenstates, will be a
subject of future investigations.

One of us (FB) would like to thank The Department of Physics of
the University of Maryland at College Park, the Department of
Physics at the University of Puebla, and the ``Cientro
Internacional de Ciencias'' at Cuernavaca where part of this work
has been done. Useful discussion with R.E.Prange, I.Guarneri,
G.Casati, V.Flambaum are also acknowledged. FMI acknowledges with
thanks financial support from CONACyT (Mexico) Grants No. 26163-E
and No. 28626-E. FB acknowledges financial support from
INFM and INFN (IS MI41).

$^{\dagger}$ On leave of absence from : Dipartimento di Matematica
e Fisica, Universit\`a Cattolica, via Trieste 17, I-25121, Brescia,
Italy.

\end{document}